\documentclass[aps,pre,onecolumn,superscriptaddress,showpacs]{revtex4-2}
\usepackage[english]{babel}
\usepackage{float}   
\usepackage{graphicx}
\usepackage{amsmath}
\usepackage{float}
\usepackage[T1]{fontenc}  
\usepackage[utf8]{inputenc}

\usepackage{color}
\usepackage[normalem]{ulem}

\newcommand{\CB}[1]{\textcolor{black}{#1}}

\begin{document}
    
\title{The simplest 2D quantum walk detects chaoticity}

\author{C. Alonso-Lobo} 
    \email{C.alonso.lobo@estudiante.upm.es}
    \affiliation{Grupo de Sistemas Complejos,
    Escuela T\'ecnica Superior de Ingener\'ia Agron\'omica, Agroambiental y de Biosistemas,
    Universidad Polit\'ecnica de Madrid,
    Avenida Puerta de Hierro 2-4,
    28040 Madrid, Spain}
\author{Gabriel G. Carlo}
\email{g.carlo@conicet.gov.ar}
\affiliation{CONICET, Comisi\'on Nacional de Energ\'ia At\'omica, 
Avenida del Libertador 8250, 1429 Buenos Aires, Argentina}
\author{F. Borondo}
\email{f.borondo@uam.es}
\affiliation{Departamento de Qu\'imica, Universidad Aut\'onoma de Madrid, 
CANTOBLANCO-28049 Madrid, Spain}

\date{\today} 

\begin{abstract}
Quantum walks 
\CB{are at present an active field of study in mathematics, with important applications in}
quantum information and statistical physics.
\CB{In this paper, we determine the influence of basic chaotic features on the walker behavior. 
For this purpose,} we consider an extremely simple model consisting of alternating 
one-dimensional walks along the two spatial coordinates in bidimensional closed domains 
(hard wall billiards). 
The chaotic or regular behavior \CB{induced by the boundary shape} 
in the deterministic classical  motion
translates into chaotic signatures for the quantized problem, 
resulting in sharp differences in the spectral statistics 
and morphology of the eigenfunctions of the quantum walker. 
\CB{Indeed, we found for the Bunimovich stadium -- a chaotic billiard -- level statistics
described by a Brody distribution with parameter $\delta \simeq 0.1$.
This indicates a weak level repulsion, and also enhanced eigenfunction localization,
with an average participation ratio ${\rm(PR)} \simeq 1150$) compared to the rectangular billiard 
(regular) case, where the average ${\rm PR} \simeq 1500$. 
Furthermore, scarring on unstable periodic orbits is observed. 
The fact that our simple model exhibits such key signatures of quantum chaos, 
e.g., non-Poissonian level statistics and scarring, that are sensitive to the underlying 
classical dynamics in the free particle billiard system is utterly surprising, 
especially when taking into account that quantum walks are diffusive models, 
which are not direct quantizations of a Hamiltonian.}
 
\end{abstract}


\maketitle

\section{Introduction} 

Quantum walks (QWs), originally proposed by Aharonov \cite{Aharonov}, 
\CB{constitute since its inception an important and active area of mathematics,
with relevant applications in} physics. 
This interest comes from their efficiency in capturing diffusive properties, 
which in turn provided many applications in quantum information \cite{QI1,QI2,Alg} 
and quantum optics \cite{QO1,QO2}. 
In particular, they are very suitable to explain and build efficient search algorithms \cite{QWbook}, 
for example in 2D grids \cite{Molfetta}, \CB {and also possibly to improve deep learning \cite{DeepLearning}.}
Moreover, QWs have been implemented in several experiments \cite{Exp1,Exp2,Exp3}.
In their simplest form, QWs are quantum counterparts of the well-known classical 
problem of the random walker on the line, taking a step to the 
right or left depending on the outcome of a coin toss. 
The straightforward quantum-mechanical counterpart can be thought of as a spin 
particle that moves to the right or left, \CB{with its}
displacement depending on its spin state, whose evolution is given by the so-called
quantum coin, the analogue of the classical coin toss.

Given the relevance of QWs in finite systems applications, studying the effects of boundaries 
is of the utmost importance \cite{Boundaries1,Boundaries2},
\CB{especially} the behavior with respect to the complexity of the
landscape in which the walker is moving. 
\CB{Relevant questions raise at this point:}
For instance, could a \textit{minimal} QW model reveal quantum signatures of 
chaos in the same way as the quantized models directly derived from the 
Hamilton's classical equations of motion do? \CB{Can the simplest 2D QW serve as a probe for classical chaoticity?}
Considering the different nature of the dynamics involved, these are challenging questions that 
we address in this work. 
In fact, QWs are not direct quantizations, in the usual sense, of deterministic classical models and \CB{may show} unexpected properties \cite{Benjamin}.
\CB{Detecting chaos with the most simple QW model not only bridges the gap between 
Hamiltonian chaos and random QWs, but it can also be related to understanding localization 
from different sources \cite{Localization}.}

In this respect, quantum chaos \cite{Haake} is \CB{the perfect arena}
where the \CB{quantum} signatures of classical chaos
have been traditionally studied. 
This area of physics \CB{has witnessed} many important  \CB{results,} being
the so-called Bohigas-Giannoni-Schmit (BGS) \cite{BGS} conjecture \CB{probably the most celebrated.}
It prescribes a random matrix behavior in quantized chaotic Hamiltonian systems. 
Notably, this subject is very active today, \CB{with many questions still open}. 
In particular, the well known level repulsion, derived from BGS, was generalized to open 
systems under the name of the Grobe-Haake-Sommers (GHS) conjecture \cite{GHS}, 
\CB{but it} has been recently challenged in \cite{Villa}. 
Another well \CB{known} phenomenon \CB{in quantum chaos is} the localization on 
marginally stable -- for example on bouncing balls and other trajectories -- and unstable orbits 
a.k.a.\ scarring; it constitutes a hallmark of quantum chaos
which always \CB{gives rise to} surprising new results \cite{Pizzi}.

\CB{Accordingly, we plan} to borrow some concepts and tools from quantum chaos 
in order to \CB{unveil} the consequences that a \CB{domain producing chaotic
dynamics} has on the QWs. 
For that purpose, we consider a very simple model consisting of a 2D QW constrained
by a hard wall. 
A deterministic classical free particle moving inside will show regular or chaotic dynamics
depending on the domain shape \cite{Haake}. 
\CB{Here,} we take the rectangle and the paradigmatic Bunimovich stadium billiards as 
benchmark examples of both behaviors, respectively. 
We conclude that a streamlined QW model consisting only 
of a single spin and alternate and independent movements along each coordinate 
perceives the different nature that these boundaries have \CB{in the dynamics}
at the classical level. 
\CB{This is reflected by the presence of scarring on unstable periodic orbits and more localization 
in the Bunimovich stadium case. 
Moreover, a non-Poissonian level statistics shows that the QW is sensitive to the 
corresponding classical dynamics of a free particle inside a billiard. 
This is a remarkable result, since  QWs are diffusive models
not directly attached to the quantization of a Hamiltonian.}

The paper is organized as follows: 
in Sec.~\ref{sec:Model} we describe our QW model, and
in Sec.~\ref{sec:Results} we show the results, where we mainly focus on the spectral 
behavior and the morphology of the spatial part of the eigenfunctions of the evolution operator. 
Finally, in Sec.~\ref{sec:Conclusions}, we present our conclusions and also outline 
some possible future developments.

\section{QW model}
\label{sec:Model}

\begin{figure}
  \includegraphics[width=.8\columnwidth]{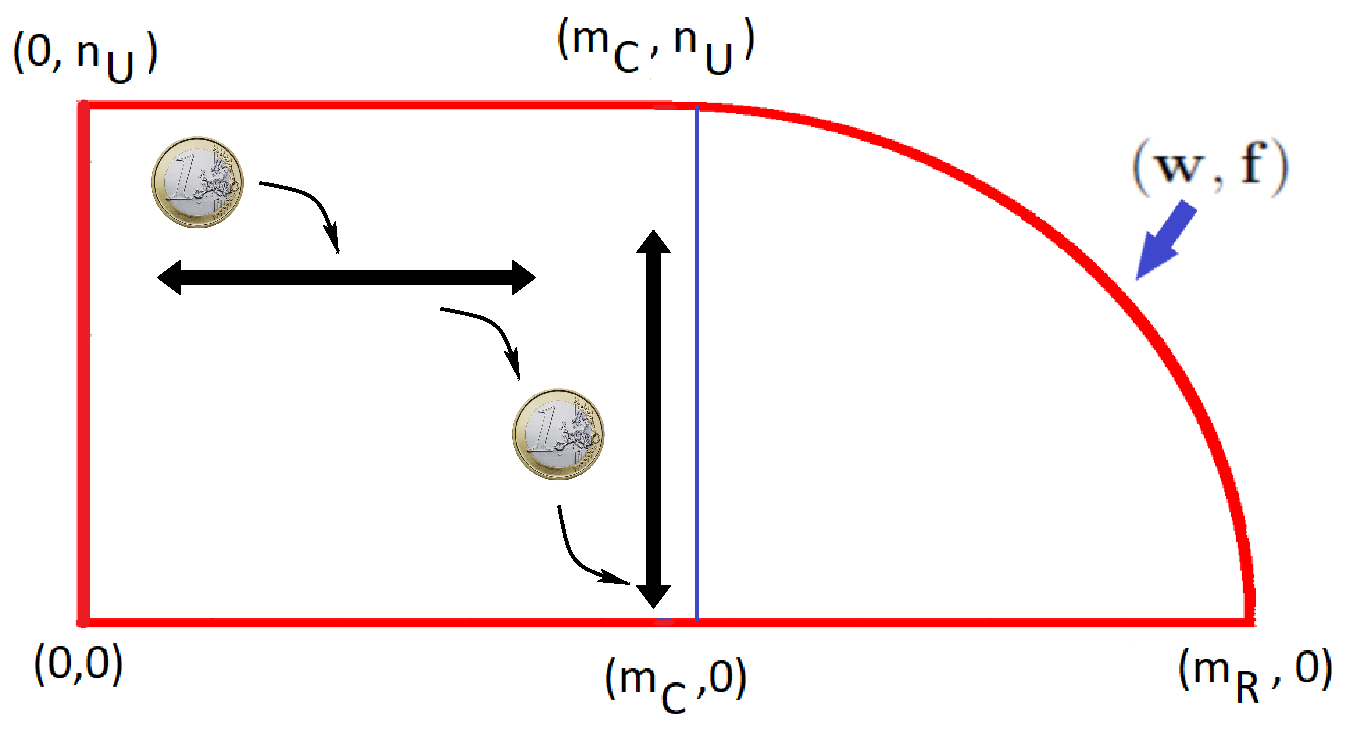} 
  \caption{(Color online) Desymmetrized Bunimovich stadium billiard. 
  Some integers $(m,n)$ specifying the particle position in a grid inside the
  billiard, and the shape functions \CB{of Eqs.~(\ref{eq:A7}) and (\ref{eq:A8})} 
  used in the evolution operator definition are shown for reference. 
  \CB{The evolution of the walker: coin 1 toss, horizontal step  (horizontal arrow), coin 2 toss, vertical step (vertical arrow), \ldots
  is schematically indicated, inside the domain.}
  See Sec.~\ref{sec:Model} for details.} 
  \label{Fig1}
\end{figure}

\CB{The simple} QW model \CB{that we have chosen to study} is defined as a 
spin $1/2$ particle moving inside a 2D billiard, whose state at any (discrete) time is 
\begin{equation}
| \Psi(t) \rangle = \sum_{m,n}  U_{m}^{n}(t) \, |m,n,u \rangle 
                          + D_{m}^{n}(t) \, |m,n,d\rangle, 
  \label{eq:Psi(t)}
\end{equation}
where $U_{m}^{n}$ and $D_{m}^{n}$ represent the probability amplitudes of the 
particle located at position $(m,n)$, with $(m,n)$ integers defining the position 
in a grid inside the billiard. 
Such particle can have either spin up, $u$, or down, $d$. 
The evolution is given by unitary operators that act on the spin space (a coin operator
reflecting the effect of a coin toss) and on the position space (a walk operator reflecting 
the step taken by the walker conditional on the spin state). 
Additionally, there is a spin flip each time the walker reaches the border of the billiard. 
\CB{(The justification of this choice is given below, at the beginning of Sec.~\ref{sec:Results}.)}
As an explicit motivation for our model, we can view the previous evolution as 
that corresponding to the motion of an electron inside a cavity. 
We can think of this model as consisting of two independent QWs, except for the fact
that both share the same spin. 
The unbounded version is usually called Alternate QW (AQW) \cite{difranco,Bru}. 

\CB{Furthermore,} we consider a bounded QW model inside the rectangular billiard and 
also the Bunimovich stadium depicted in Fig.~\ref{Fig1}.
For comparison purposes, the rectangular billiard is
taken as the rectangle in which the Bunimovich stadium is contained,
both having a horizontal length two times the vertical one. 
Recall that the classical dynamics in the rectangular domain is regular.
\begin{figure*}[t!]
  \includegraphics[width=0.8\textwidth]{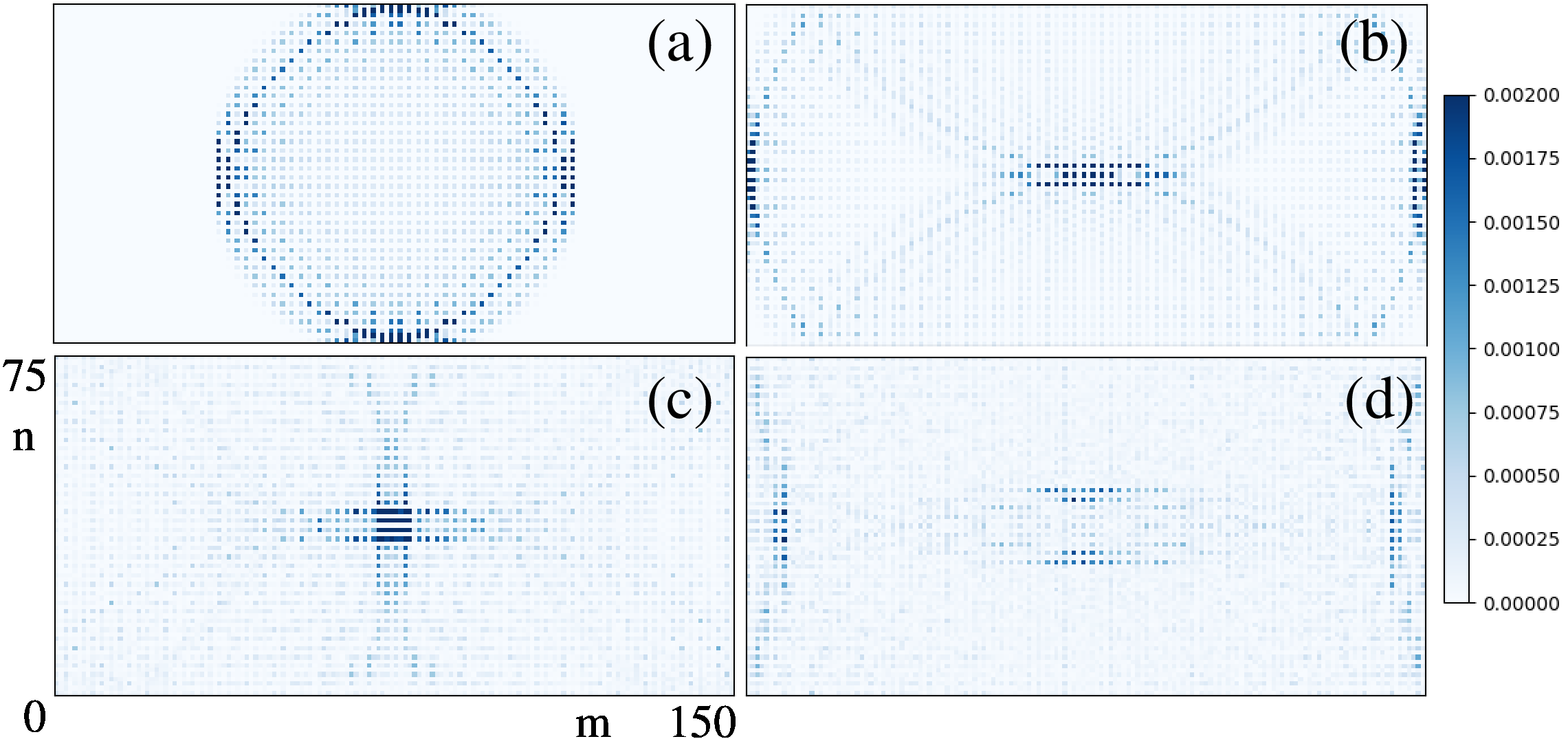}
  \caption{(Color online) Time evolution of the central position eigenstate 
  (see main text for details) for $t$ = 38 (a), 76 (b), 152 (c), 232 (d) in the rectangular billiard. 
  A $(m_R,n_U)=(150, 75)$ position grid,  and $\alpha=\beta=\pi/4$ have been used in 
  the calculations.
  \CB{The color scale bar at the right represents the probability of the wavefunction.} }
  \label{Fig2}
\end{figure*} 
%
\begin{figure*}
  \includegraphics[width=0.8\textwidth]{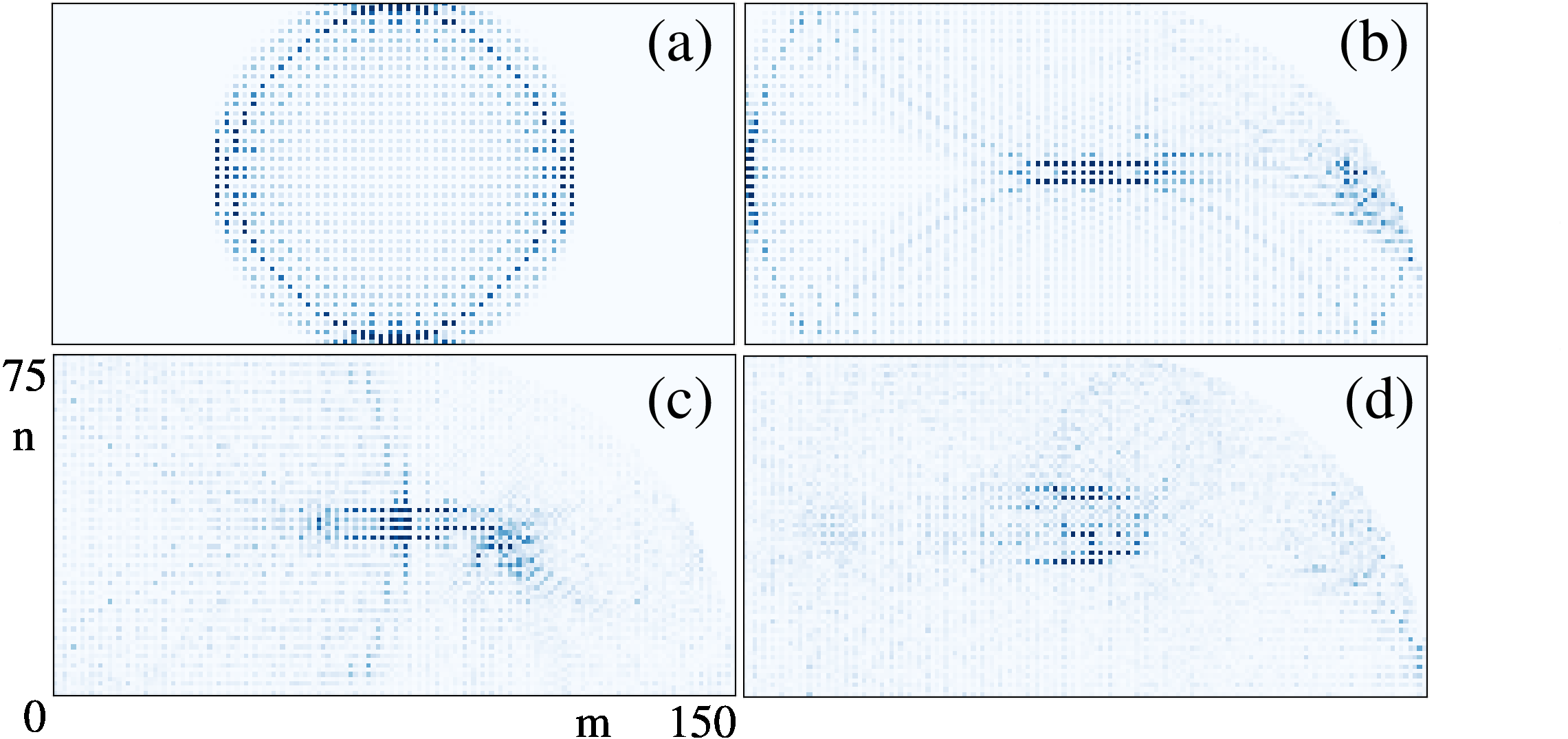}
  \caption{(Color online) Same as Fig.~\ref{Fig2} for the Bunimovich stadium billiard.}
    \label{Fig3}
\end{figure*}  
\CB{The evolution of our walker can be summarized \CB{algorithmically} by the following 
iterative sequence:
\\ \indent (Quantum) Coin toss $\rightarrow$ 
\\ \indent Horizontal step (with eventual boundary spin flip) $\rightarrow$ 
\\ \indent (Quantum) Coin toss $\rightarrow$
\\ \indent Vertical walk (with eventual boundary spin flip) $\rightarrow$
\\ (see inner part of the Bunimovich stadium in Fig.~\ref{Fig1}). 
The corresponding operator for one time step is given by 
$\hat{Q_w}=\hat{W_n} \hat{C}_2 \hat{W_m} \hat{C}_1$, 
where $\hat{W_m}$ and $\hat{W_n}$ are horizontal and vertical displacement operators
respectively, which include reflection at the boundaries 
(by means of the previously mentioned spin flip), 
and $\hat{C}_2$ and $\hat{C}_1$ are the coin operators.
See more details in the first part of the Appendix.}

\CB{To conclude this section, we discuss the time evolution under this dynamics 
of a representative position eigenstate on the two spatial domains (rectangular and Bunimovich
stadium billiards) considered in our calculation.}
In Fig.~\ref{Fig2} we show some snapshots of the evolution of the corresponding probability 
density distribution for the rectangular billiard case. 
\CB{A grid of size} $m_{R}=150$ and $n_{U}=75$, and an initial eigenstate 
position at the center ($m=75,n=37$) \CB{is taken in our calculations}; 
for spin up the amplitude is $1/\sqrt{2}$ while for spin down is $i/\sqrt{2}$. 
\CB{Four representative snapshot times have been chosen so as to (approximately) 
cover the lapse of time taking the `border' of the wavefunction to return the to center of the 
billiard, after bouncing once at the walls.} 
\
\CB{Similar results are presented} for the Bunimovich stadium  in Fig.~\ref{Fig3}. 
\CB{The results clearly show} that before \CB{the time corresponding to the} bounce 
on the circular part of the Bunimovich stadium both probability distributions coincide,  
as expected. 
\CB{However,} after the main reflection on the circle at $t=76$ they become markedly different,
this showing the regularity and strong symmetry existing in the rectangular billiard 
(aside of minor details due to the grid dimensions and coin shape) and the lack of them in the
stadium. 
In fact, the probability peak at the center \CB{in the case} of the rectangular domain is gradually 
destroyed in the Bunimovich case for later times as can be checked by comparing the lower panels of both figures.

\begin{figure*}
  \includegraphics[width=0.8\textwidth]{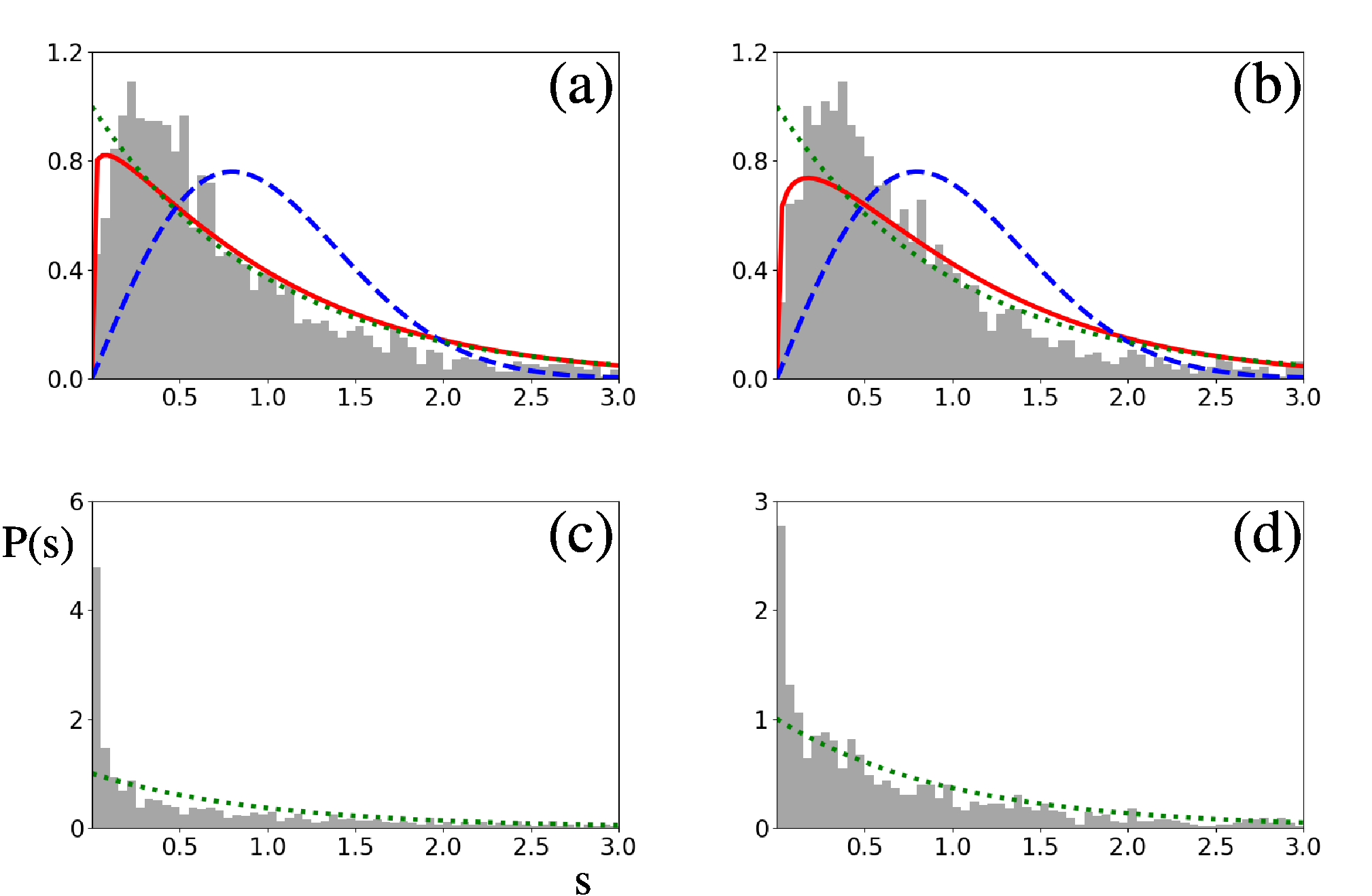} 
  \caption{(Color online) Unfolded level spacing distributions $P(s)$ vs.~$s$
  for the Bunimovich stadium billiard with coin angles 
  $\alpha = \beta = \pi / 4$ (a) and
  $\alpha = \pi / 4$ $\beta = \pi / 3$ (b),
  and for the rectangular billiard with coin angles 
  $\alpha = \beta = \pi / 4$ (c) and
  $\alpha = \pi / 4$ $\beta = \pi / 3$ (d). 
  \CB{The corresponding Wigner surmise $P_{\rm W}$ of Eq.~(\ref{eq:Wigner}), 
  best fitting Brody distribution $P_{\rm B}$ of Eq.~(\ref{eq:Brody})
  (with the parameters and errors reported in Table~\ref{tab:freq}), and 
  the Poisson distributions of Eq.~(\ref{eq:Poisson}),} 
  are also shown, \CB{for comparison,} 
  as blue (black) dashed lines, red (gray) solid lines and green (light gray) dotted lines,
  respectively. 
  }
\label{Fig4}
\end{figure*}

\section{Chaotic signatures: spectral behavior and morphology of the eigenfunctions}
\label{sec:Results}

\CB{An important point concerning our QW model is worth discussing here.}
An unbounded one-dimensional QW is translationally invariant, and as a consequence 
a simple \textit{Dispersion Relation} (DR) can be obtained. 
A similar situation arises for the unbounded AQW in two dimensions \cite{Roldan}, 
leading to a generalized DR, which is a straightforward extension of the 1D solution.
\CB{In the case of} a cylinder, i.e.~a compact domain with periodic boundary conditions 
in one direction, the same situation effectively happens, and a DR has been derived for 
this case \cite{Bru2}. 
On the other hand,
\CB{a very interesting class of} billiards consists of a completely bounded domain 
with reflective boundaries \CB{as the case considered in this work.}
\CB{In this case, the effect of the boundaries in} QWs can be implemented simply by means 
of spin flips. 

\subsection{Spectral statistics}

Since in our case there is no analytical DR, in what follows we will proceed using numerical 
explorations of the behavior of our system, considering spectral statistics and the 
morphology of the corresponding eigenfunctions. 

\CB{For this purpose,} we directly diagonalize the evolution operator for one time step, 
i.e.~$\hat{Q_w}$, in the spin-position basis set, and then study its (unfolded) spectrum, 
\CB{for cases considered here.} 
To speed up the computations \CB{(which are heavy)}, we here take a \CB{(smaller)} grid of size 
$(m_R, n_U)=(50, 25)$.
(Notice that this grid implies a smaller position basis for the Bunimovich stadium 
\CB{than for the rectangular billiard.})
\CB{Moreover, though this choice generates finite-size effects, we have found that it represents a fair
compromise between computational cost and physical significance.}
It is also important to emphasize that desymmetrizing the Bunimovich stadium is a customary 
procedure to avoid unwanted symmetries, and then unveil the eventual Wigner surmise 
(or other statistics in different systems) for the eigenvalues associated to the corresponding
Hamiltonian. 
Failing to do so, allows the \CB{unwanted} overlapping of uncorrelated spectra corresponding 
to different symmetry classes. 
Keeping our line of reasoning, we will proceed in this way below.

We show in Fig.~\ref{Fig4} the spectral statistics for our model, where the eigenphases 
(i.e.~the phases of the complex unimodular eigenvalues) of the evolution operator have been
considered \CB{in the construction} of the histograms, $P(s)$, for the unfolded level spacing $s$. 
\CB{Indeed,} after ordering the eigenphases between $0$ and $2\pi$, the unfolding 
\CB{of these values} is performed by simply dividing the distances between nearest neighbors 
by the \CB{corresponding} mean (arithmetic) distance.
\CB{Obviously, more sophisticated unfolding procedures, such as for example fitting a 
polynomial to the cumulative level density,  can be used. 
However, for our purposes, i.e.~a straightforward detection of spectral behavior differences, 
the simple method used is plenty enough.} 
The results for two different pairs of $\alpha$ and $\beta$ values \CB{(coin phases)} are shown
in the different panels \CB{of Fig.~\ref{Fig4}}; 
\CB{being equal (symmetrical coins) in (a) and (c) , 
but different (asymmetrical coins) in (b) and (d), see caption for details.} 
Although not directly applicable, in principle, to our QW scenario, 
we plot in each panel \CB{together with the histograms for comparison,}  
the so-called Wigner surmise \cite{Wigner} 
%
\begin{equation} 
  P_{{\rm W}}(s) = \frac{\pi}{ 2}  \ s \  \exp{\left(\frac{ - \pi \ s ^{2}}{4}\right)},
   \label{eq:Wigner}
\end{equation}
as a blue (black) dashed line.
\CB{This surmise is known to} describe the spectral behavior of chaotic Hamiltonians.
In the other extreme of level statistics, we have that for regular systems, spectral 
spacings are known to satisfy the Poisson distribution, given by 
\begin{equation}
  P_{{\rm P}}(s) = \exp{(-s)}.
  \label{eq:Poisson}
\end{equation}
Given that the QW dynamics is different from the one corresponding to
the quantum Hamiltonian of a free particle inside a billiard,
deviations from these two analytical expressions are expected in our results. 
Hence, we have also fitted the data for the Bunimovich case, 
using least squares and the root mean square \CB{(RMS)} error, to the Brody distribution \cite{Brody},
a very simple function that interpolates between Poisson and the Wigner surmise,
given by the expression
\begin{equation} 
P_{{\rm B}}(s) = a \ s^{\delta} \  \exp{(-b\ s^{\delta+1})},
  \label{eq:Brody} 
\end{equation}
where $a=(\delta+1)b$, $b={\Gamma((\delta+2)/(\delta+1))}^{\delta+1}$, 
being $\delta$ a fitting parameter between $0$ (Poisson distribution) and $1$ (Wigner surmise). 
The corresponding results are displayed by means of red (gray) solid lines
in Fig.~\ref{Fig4} (a) and (b), and the fitting values are reported in Table~\ref{tab:freq}.
We notice \CB{by direct eye examination} that our results behave very differently from those of 
the Wigner surmise, \CB{something also ascertained by the values of $\delta=0.07$ and $0.15$ 
and the fact that the error is less than 0.5 (see Table~\ref{tab:freq}). }
By inspection, it is also evident that the Poisson ($\delta=0$) curve is closer to the histogram bars, 
though there is \CB{some relevant} level repulsion. 
\CB{This is a remarkable result in that it suggests a novel spectral behavior, 
not previously described, to the best of our knowledge, in the literature.}
\CB{The reason for this can be that} more symmetries besides those already broken exist 
(almost immune to the asymmetrical coins choice) and uncorrelate the spectrum leading 
to this peculiar Poisson-like statistics with level repulsion. 
\CB{In the future, we will investigate a model that can explain this kind of spectral behavior, 
perhaps related to the Anderson transition or other properties of disordered systems.}

\begin{table}
\caption{\CB{Best fitting Brody parameter and} errors \CB{on the Brody and Wigner spectral 
distributions for the results of our QW model} in the Bunimovich stadium billiard.} 
\label{tab:freq}
\begin{tabular}{ccccc}
\hline\hline
\multicolumn{2}{c}{Coins}  & & \multicolumn{2}{c}{RMS Error} \\
$\alpha$ & $\beta$ & $\delta$  & $P_{\rm B}$ & $P_{\rm W}$ \\
\hline
$\pi / 4$ & $\pi / 4$ & 0.07 & 0.044 & 0.131 \\
$\pi / 4$ & $\pi / 3$ & 0.15 & 0.069 & 0.154 \\
\hline 
\end{tabular}
\end{table}

In the case of the rectangular billiard, whose associated Hamiltonian is regular, 
we have evaluated the error of the Poisson distribution
(see Fig.~\ref{Fig4} (c) and (d)). \CB{The results are reported in Table~\ref{tab:freq2}.}
\begin{table}[b]
\caption{Poisson distribution errors 
for the QW in the rectangular billiard} 
\label{tab:freq2}
\begin{tabular}{cccc}
\hline \hline
\multicolumn{2}{c}{Coins} & \multicolumn{2}{c}{RMS Error} \\
$\alpha$ & $\beta$ & $P_{\rm P}$  \\
\hline
$\pi / 4$ & $\pi / 4 $ & 0.140 \\
$\pi / 4$ & $\pi / 3 $ & 0.090 \\
\hline 
\end{tabular}
\end{table}
As can be clearly seen, the behavior is not strictly Poissonian, although it resembles it with the 
exception of the very large first bar, something which signals a particularly high lack of level repulsion, which again is almost immune to the asymmetrical coins choice. 

\CB{Summarizing,} the shape of the histograms in the Bunimovich and the rectangular billiards 
strongly differ in the small spacing region. 
Though the QW dynamics is clearly different from that of a quantum free particle moving inside 
a hard wall billiard, the statistical behavior of the spectra in these two cases allows to distinguish 
between them. 

\CB{At this point, another interesting question arises:}
Can we also identify other differences of this kind, reminiscent of the ones found in the Hamiltonian
dynamics, at the system eigenfunction level?

\subsection{Eigenfunctions morphology}

\begin{figure*}
  \includegraphics[width=.8\textwidth]{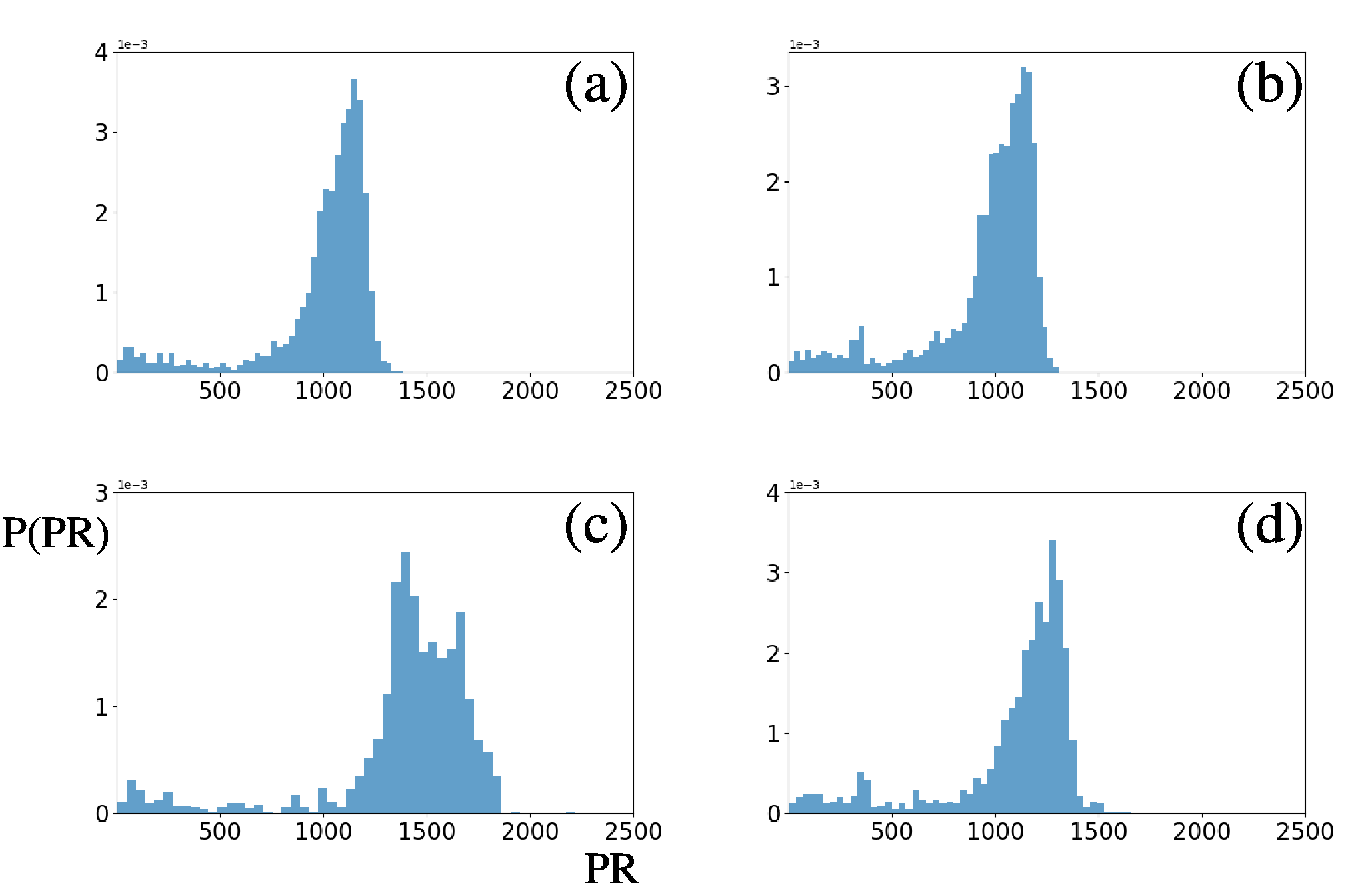} 
  \caption{(Color online) \CB{Distribution of the participation ratios of Eq.~(\ref{eq:PPR}),}
  $P(\rm{PR})$ vs.~PR, for the Bunimovich stadium with $\alpha =  \beta = \pi \ / 4$ (a), and 
  $\alpha =  \pi \ / 4$, $\beta = \pi \ / 3$ (b); and  
  the rectangular billiard with $\alpha =  \beta = \pi \ / 4$ (c) and 
  $\alpha =  \pi \ / 4$, $\beta = \pi \ / 3$ (d).
  } 
  \label{Fig5}
\end{figure*}

To answer this question, we resort to analyzing the morphology of the eigenfunctions. 
For that purpose, we use the \textit{Participation Ratio} (PR), a commonly used measure of localization 
in quantum chaos \cite{Haake}. 
The PR is calculated by first normalizing the eigenfunctions $|\Phi\rangle$, 
and then evaluating 
\begin{equation}
{\rm PR}=\left(\sum_{m,n} |{U_{\Phi}}_{m}^{n}|^{4}+|{D_{\Phi}}_{m}^{n}|^{4}\right)^{-1},
\label{eq:PPR}
\end{equation}
where ${U_{\Phi}}_{m}^{n}$ and ${D_{\Phi}}_{m}^{n}$ are the components of $|\Phi\rangle$). 
For example, in the \CB{case of the rectangular billiard}, the values of the PR range from $1$ to 
$2 m_R n_U$ ($2 \times 50 \times 25$), this roughly indicating how many position basis elements 
participate into a given eigenfunction (besides the factor $2$ coming from the spin part of the Hilbert space). 
As a consequence, larger PR values indicate less localized states in this basis. 
In Fig.~\ref{Fig5} we show the histogram corresponding to $P(\rm{PR})$ vs.~PR.
It can be directly observed that localization on $\sim$1150 basis elements is the approximate 
typical behavior for the Bunimovich stadium, while \CB{this value amounts} to $\sim$1500 and 1400 
for the \CB{case of the rectangular billiard}. 
\CB{Recall this is partly} due to the different effective dimensions of the position basis in the two cases; 
being the Bunimovich stadium basis size approximately $0.91$ times that of the rectangular billiard. 
This is around half of the bases sizes (again, \CB{take into account} the spin space dimension). 
\CB{However, not only are the PRs in the Bunimovich stadium billiard more localized} than in 
the rectangle, on average, \CB{but also, their} distribution is more biased towards 
lower PR values in a meaningful way (PR approximately $\in [600,800]$), 
range where those for the rectangular billiard have smaller values). 
In what follows, \CB{we are going to deepen into these features, 
by analyzing some representative cases.}

\begin{figure*}
  \includegraphics[width=0.8\textwidth]{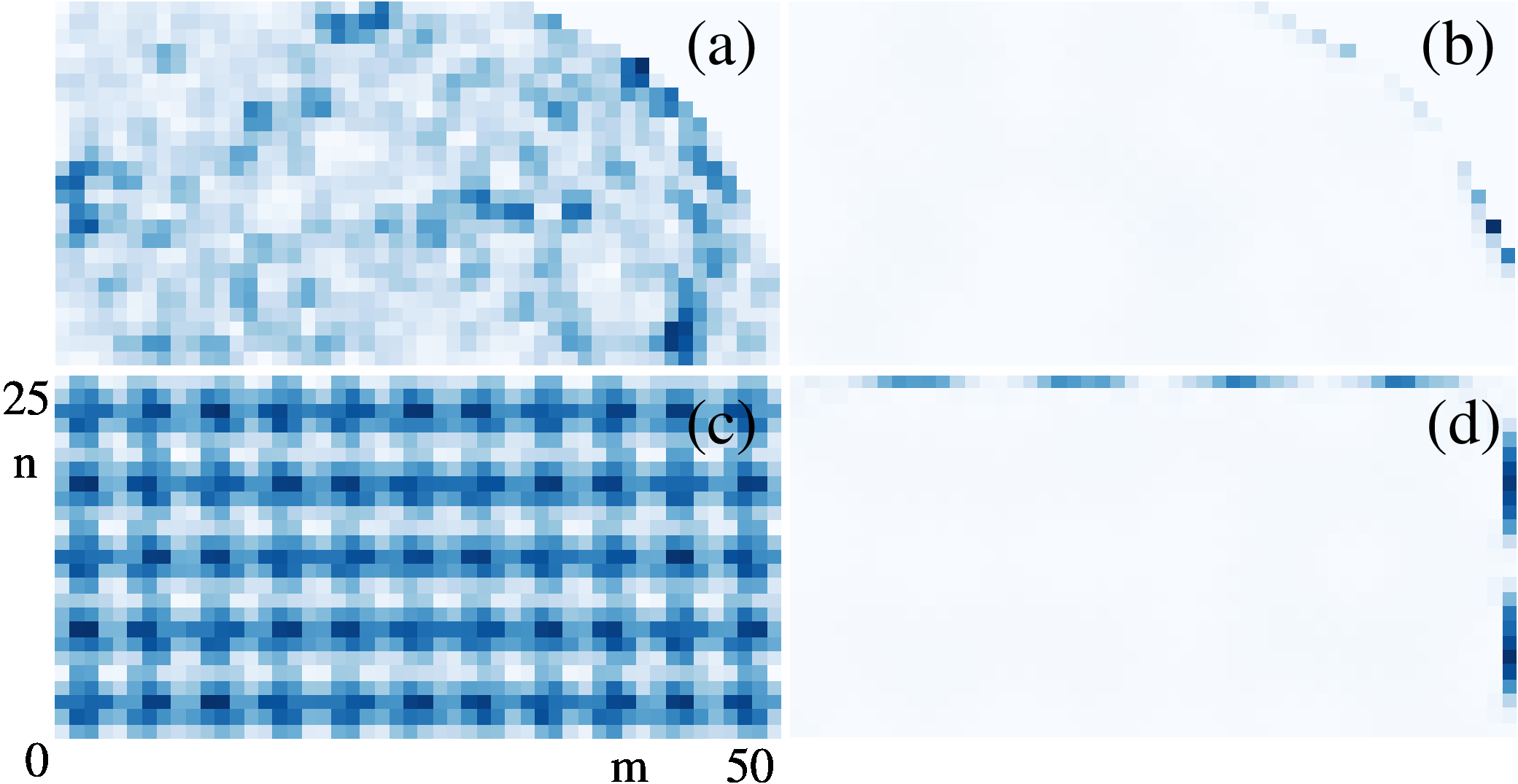} 
  \caption{(Color online) \CB{Eigenfunctions of the QW evolution operator for:}
  (upper row) the Bunimovich stadium with PR=$1152.83$, eigenphase=$-1.0391$ (a), and
   PR=$116.58$, eigenphase=$1.3166$ (b); and 
   (lower row) the rectangular billiard with PR=$1808.39$, eigenphase=$1.6833$ (c) and
    PR=$116.2$, eigenphase=$-1.9794$ (d). 
    In all cases $\alpha = \beta = \pi \ / 4$ \CB{(symmetrical coins)}.
    \CB{The same color scale of Fig.~\ref{Fig2} is used here.}}
  \label{Fig6}
\end{figure*}

We next display and analyze some examples of eigenfunctions that help to understand the 
behavior of  \CB{wavefunction localization in our QW model}.
\CB{States are labeled with their corresponding values of the PR and eigenphase in the 
figure captions}. 
In Fig.~\ref{Fig6}, some examples of the most delocalized \CB{(left column)}
and most localized \CB{(right column)} eigenfunctions
in the Bunimovich stadium (upper row) and in the rectangular billiard (lower row) are shown; 
both correspond to QWs with symmetrical coins. 
In it, we notice that the two delocalized states (left column) are very different, 
with \CB{that corresponding to the Bunimovich stadium showing a 
more \textit{irregular} (or \textit{chaotic} in the sense of the quantum chaos theory \cite{Haake}),
and that for the rectangular billiard} looking markedly more regular. 
For these maximally delocalized states, the analogy with the Hamiltonian system 
\CB{is obvious}, and the different nature of the dynamics clearly manifests itself.
\CB{However, the situation is completely different for the most localized states presented
in the right column.} 
As a matter of fact,\CB{here we observe} eigenfunctions extremely peaked near the boundaries,
which are not usually found in quantum billiards. 
\CB{In our opinion the nature and origin of this states} deserve further study, 
\CB{which is beyond the scope of the present work}.
%
\begin{figure*}
  \includegraphics[width=0.8\textwidth]{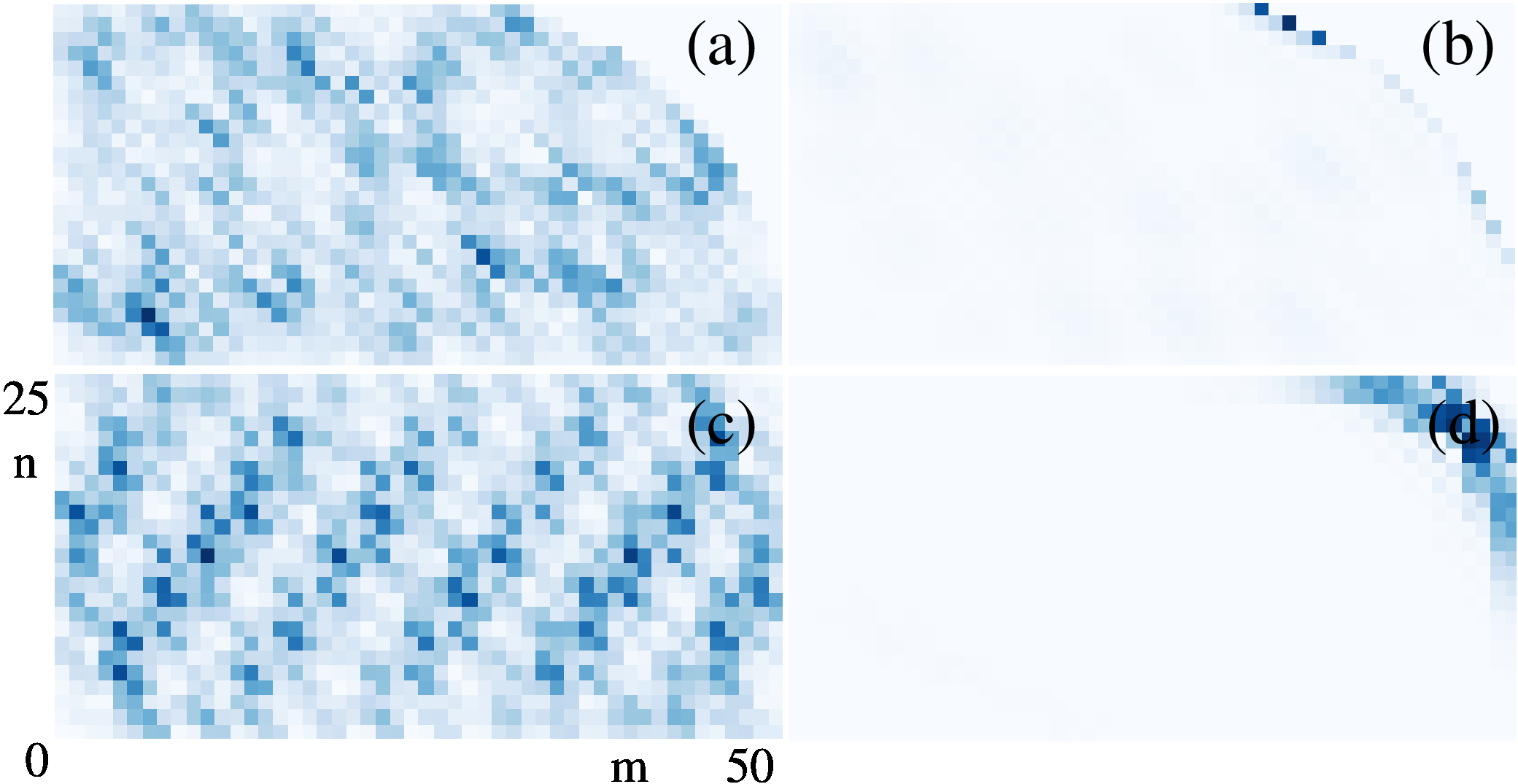} 
  \caption{(Color online) 
  \CB{Same as Fig.~\ref{Fig6} for the case of asymmetrical coins.
  Here we have: PR=$1171.65$, eigenphase=$2.9584$ (a);
  PR=$115.09$, eigenphase=$1.4287$ (b); PR=$1286.86$, eigenphase=$-0.4716$ (c); 
  PR=$102.99$, eigenphase=$-0.8186$ (d), and
  $\alpha = \pi \ / 4$ and $\beta = \pi \ / 3$.}}
  \label{Fig7}
\end{figure*}

\CB{The case of asymmetrical coins is considered} in Fig.~\ref{Fig7}. 
In this case both systems apparently behave in an irregular fashion, 
\CB{but a closer examination reveals that} the only change from the previous results is 
the tilting induced by the  asymmetry of the coins,
that clearly bias one of the 1D translations with respect to the other.
\CB{This results in a probability density distributed more along diagonal lines.} 
A remark about the effect of not using an unbiased Hadamard-like coin is \CB{in order} here. 
When considering $\beta \in (\pi/4;\pi/2]$ we anticorrelate more and more the horizontal 
and vertical walk directions, thus reaching a displacement along the main diagonal of the 
position domain at $\beta=\pi/2$. 
The same happens for  $\beta \in [0;\pi/4)$, but in this case the two walking directions 
become more and more correlated towards $\beta=0$, where a displacement along the other diagonal happens. 
In any case, the differences between regular and irregular behavior due to the shape of the
boundaries in the corresponding Hamiltonian dynamics can still be detected given 
that we are not in these limits \CB{when taking $\beta = \pi \ / 3$; 
this value represents well the typical biased but still not completely correlated scenario.}

\begin{figure*}
  \includegraphics[width=0.8\textwidth]{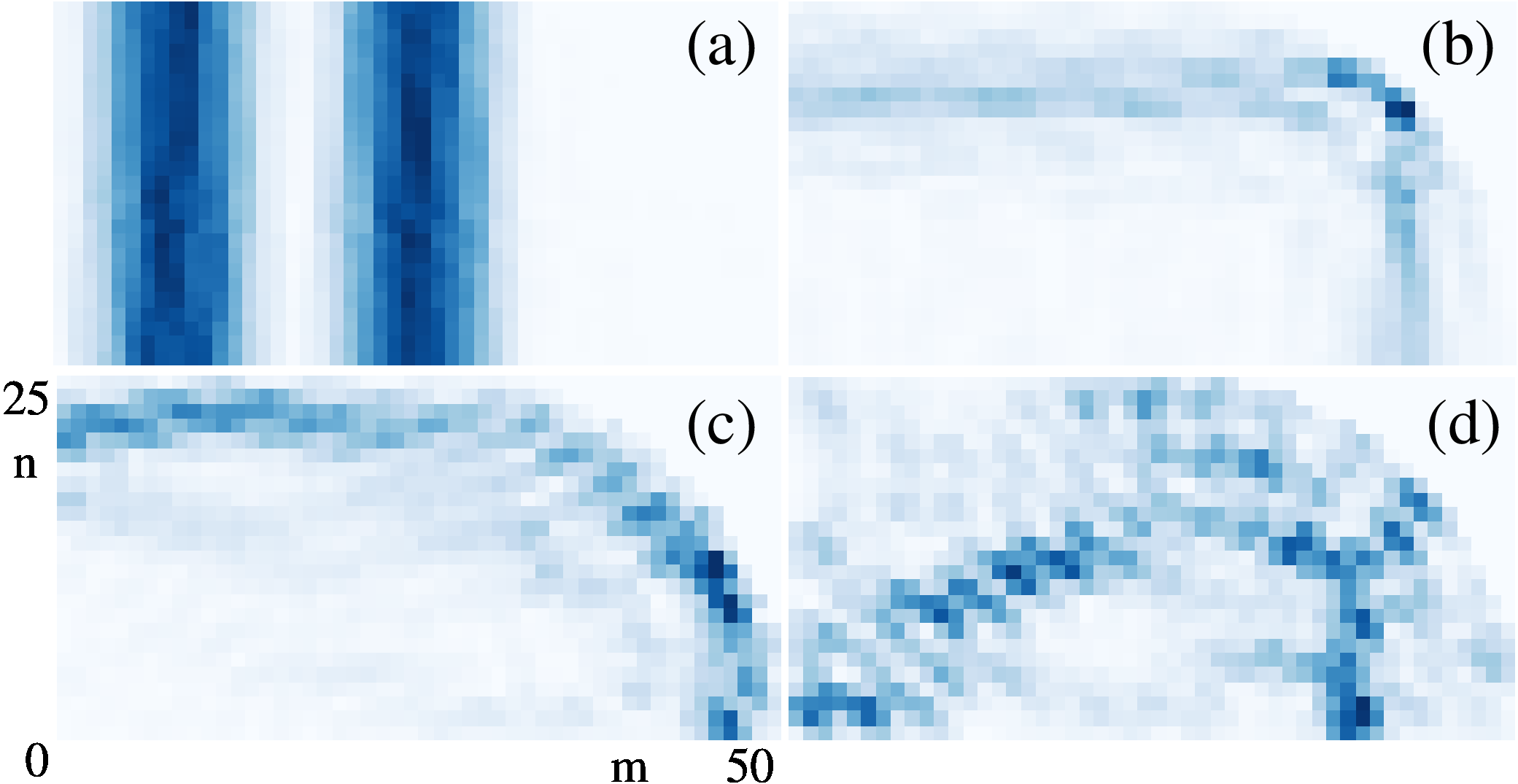} 
  \caption{(Color online) Bouncing ball and scarred \CB{eigenfunctions of the QW evolution 
  operator for the Bunimovich stadium,} with
   PR=$737.00$, eigenphase=$2.2945$ (a);
  PR=$673.79$, eigenphase=$2.3170$ (b);
  PR=$737.36$, eigenphase=$2.0209$ (c); and 
  PR=$921.34$, eigenphase=$-0.2780$ (d). 
  In all cases $\alpha = \beta = \pi \ / 4$.
  \CB{The same color scale as in Fig.~\ref{Fig2} is used.}}
  \label{Fig8}
\end{figure*}

\begin{figure*}
  \includegraphics[width=0.8\textwidth]{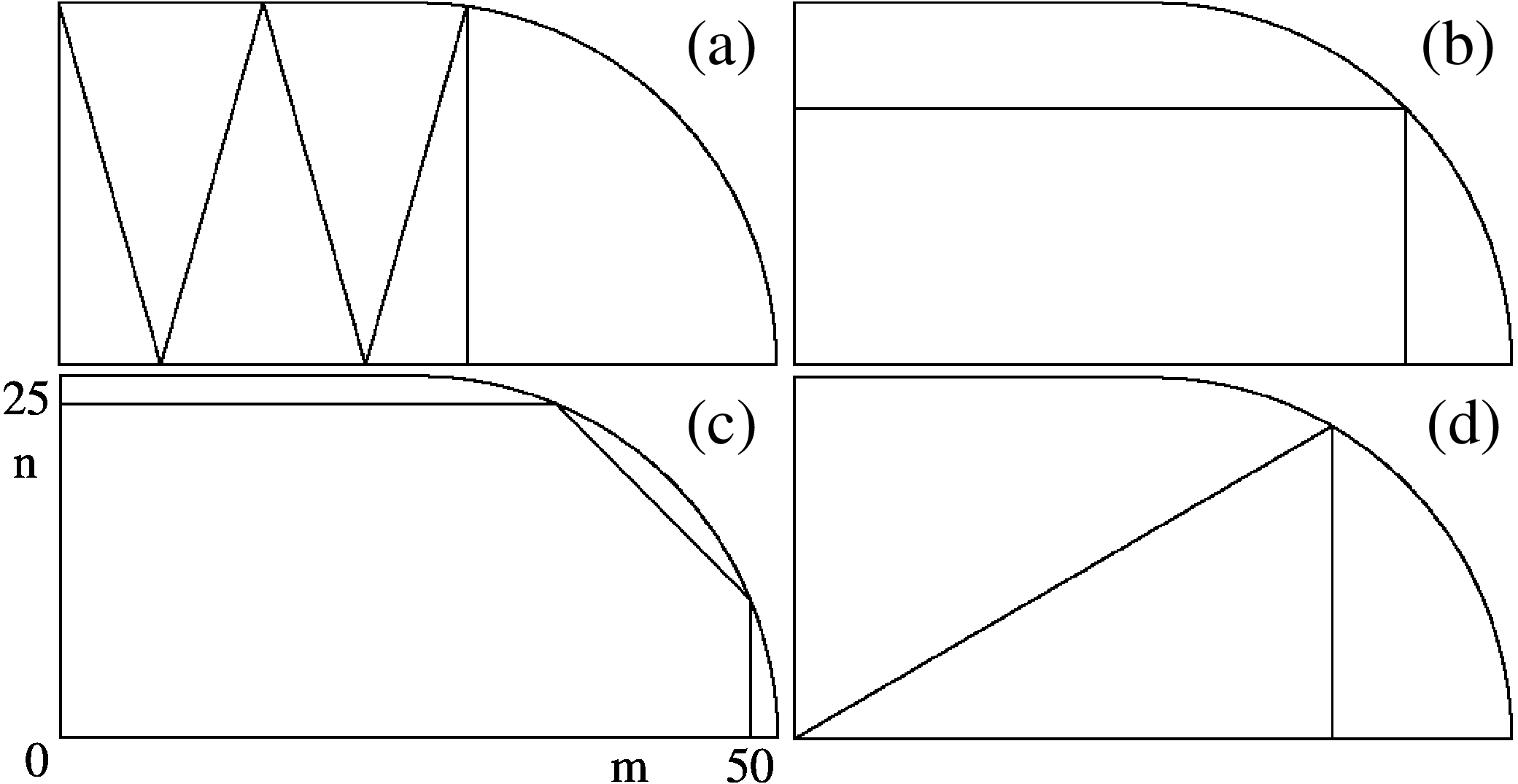}
  \caption{\CB{Classical periodic orbits of the desymmetrized Bunimovich
  stadium: (a) bouncing ball; (b)
rectangular; (c) whispering gallery; and (d) bow-tie used in the construction of the scar functions with which we overlap the bouncing ball
  and scarred eigenfunctions of Fig.~\ref{Fig8}.
  See text for details.}}
  \label{Fig9}
\end{figure*}

More interestingly, in the case of the Bunimovich stadium billiard, 
we have also found localization on structures similar to those found in the corresponding 
Hamiltonian system, \CB{(by using an automated search for intermediate PR states followed by visual
inspection)}. 
\CB{This effect can} be responsible for the greater localization found in this case 
(notice their typical PR values). 
In fact, we have found bouncing ball states \cite{Haake} that closely resemble those which are
ubiquitous in the quantum chaos literature \cite{ours,Selinummi}. 
\CB{A representative example is} shown in Fig.~\ref{Fig8} (a) for the symmetrical coins case
In the standard quantum chaotic model, this family of eigenfunctions are localized on marginally stable orbits that form a continuous family. 
In the QW scenario \CB{their manifestation} is remarkable,
having in mind that a purely diffusive behavior is expected. 
The rest panels of Fig.~\ref{Fig8} correspond to scarred states \cite{Haake} on short periodic 
orbits (POs) of the Hamiltonian classical system, namely the ``rectangle'' orbit in (b), 
a whispering gallery mode (which is a special case similar but not equal to the bouncing ball family) in (c) and a scar on the ``bow tie'' trajectory in (d).
\CB{See also a representation of these POs en Fig.~\ref{Fig9}.}

\CB{Scarring is a well described phenomenon in quantum chaos.
The term scar was first coined by Heller \cite{Heller,Revuelta} to define an unexpected 
localization along unstable POs of the probability density of certain eigenstates of
Hamiltonian systems, whose classical behavior was chaotic.
Since then it became one of the cornerstones in quantum chaos, since 
demonstrated the existence of quantum correspondence of classical chaos.}

\CB{Next, making a quantitative evaluation of scarring effect present in the eigenfunctions
of our QW model evolution operator is in order.
For this purpose, we resort to the method of
semiclassical construction of resonant (or scar) functions along POs.
This method will be briefly described next, but we refer the reader to reference
\cite{ours} and references therein for more details.
These resonances can essentially be seen as the product of a plane wave in the direction 
along a given scarring PO, using the semiclassical approximation for the unidimensional 
motion along the orbit, and a Gaussian wave packet in the transverse coordinate, 
which follows a dynamics without dilation–contraction along the unstable and stable
manifolds of the unstable trajectory. 
The wavenumber $k$ is approximately given by the Bohr-Sommerfeld quantization rule 
$k L=2 \pi n$, $L$ being the length of the PO. 
(For simplicity we ignore here, but not in the calculations, 
topological contributions \cite{Haake}, such as Maslov indices and boundary conditions, 
that guarantees continuity along the PO.)
In Fig.~\ref{Fig9} we show the relevant POs needed to evaluate the localization of the 
eigenfunctions in Fig.~\ref{Fig8}. 
We notice that the POs in Fig.~\ref{Fig9}(a) and (c) are suitable to build scar functions that 
well approximate bouncing ball and whispering gallery kind of eigenfunctions, 
but also that they are not the only possible choice since they originate in singular families of POs.}

\begin{figure*}
  \includegraphics[width=0.8\textwidth]{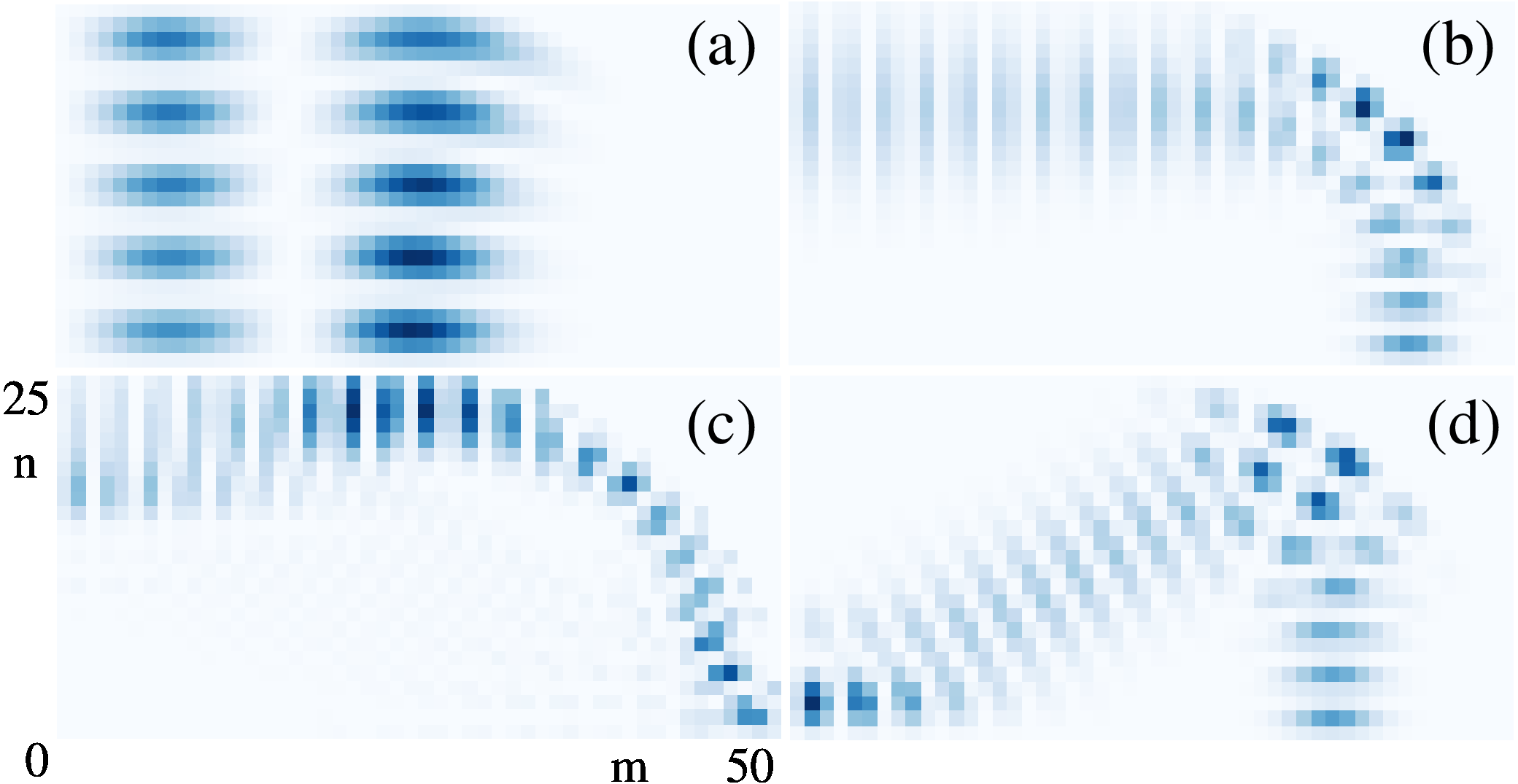}
  \caption{\CB{(Color online) Bouncing ball and scar eigenfunctions of a quantum particle
  on the Bunimovich stadium for values of the corresponding wavenumbers: 
  $k$=16.5027 (a), 29.2790 (b), 29.7930 (c), and 27.8116 (d).
  The corresponding overlaps with the QW eigenfunctions of Fig.~\ref{Fig8} are:
  0.7948, 0.6471, 0.5877, and 0.6271, respectively.
  The same color scale of Fig.~\ref{Fig2}has been used.}}
  \label{Fig10}
\end{figure*}

\CB{To ascertain the scarring character of the eigenstates in Fig.~\ref{Fig8},
we have evaluated the overlap between the probability distributions corresponding 
to the scar functions computed on these POs (that are shown in Fig.~\ref{Fig10}) 
and those corresponding to the bouncing ball and scarred eigenfunctions in Fig.~\ref{Fig8}. 
The results are given in the caption of Fig.~\ref{Fig10}. 
It is interesting to note that it is possible to obtain a $60\%$ or more of the eigenfunctions 
with such a simple and semiclassical approximation, and this figures goes to almost $80\%$ 
in the case of the bouncing ball but due to the extreme localization on a given region 
of the whole domain existing in this case.}

\begin{figure*}
  \includegraphics[width=0.8\textwidth]{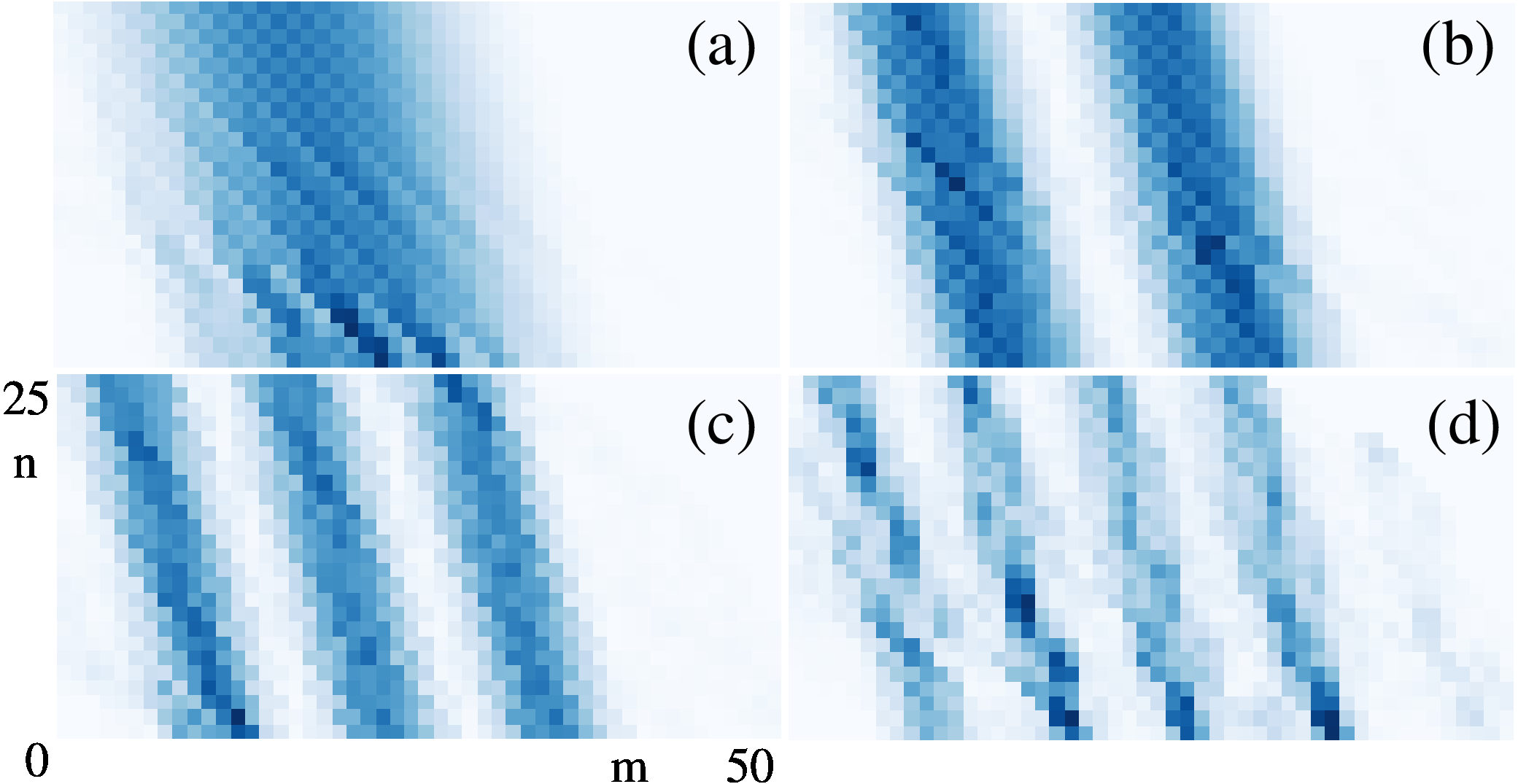}
  \caption{(Color online) Bouncing ball eigenfunctions of \CB{the QW evolution operator} for
  the Bunimovich stadium. 
  \CB{Here PR=$718.73$, eigenphase=$-0.7249$ (a); 
  PR=$690.18$, eigenphase=$-0.7257$ (b); 
  PR=$678.01$, eigenphase=$-0.8440$ (c);
  PR=$873.06$, eigenphase=$-0.6140$ (d); and
  $\alpha = \pi \ / 4$ and $\beta = \pi \ / 3$.
  The same color scale as in Fig.~\ref{Fig2} is used.}}
  \label{Fig11}
\end{figure*}

\CB{Finally, it is noticed that the typical sequence of horizontal excitations 
of the bouncing ball states is also present in the QW for both, symmetric and
asymmetric, coin cases. 
The only difference happens when considering the asymmetrical coin
which just tilt them. 
One example of this sequence is shown in Fig.~\ref{Fig11}.}

\CB{Localization on POs of the analogue classical Hamiltonian system by a diffusive 
quantum model that is not a direct quantization of it, is a very striking result. 
We think that our very simple diffusion model along orthogonal directions in configuration 
space is enough to mimic the evolution provided by the Helmholtz equation inside a billiard. 
But if all the standard scarring can be reproduced with it, this still remains an open question
that certainly deserves further investigation.}
\CB{Finally, we present Table \ref{tab:summary} which summarize our main obtained QW 
results, especially pointing out the differences found between the regular and the chaotic
billiards considered in this work.}

\begin{table}
\caption{\CB{Summary of results for both billiard shapes considered in this paper.}}
\label{tab:summary}
\begin{tabular}{lcc}
\hline\hline
 & \CB{Rectangular} &  \CB{Bunimovich} \\
\hline
\CB{Character} & \CB{regular} & \CB{chaotic} \\
\CB{Poisson vs.\ Brody}  &    \CB{RMSE = 0.09 $\sim$ 0.14} &  \CB{$\delta=0.07\sim0.15$} \\ 
\CB{Average PR} & \CB{$1500$} & \CB{$1150$} \\
\CB{Scarring} & \CB{no} & \CB{yes} \\
\hline
\end{tabular}
\end{table}

\section{Conclusions}
\label{sec:Conclusions} 

In these paper, we have studied the properties of a QW taking place inside compact 
bidimensional domains with different boundary shapes, which are characterized by regular and irregular behaviors of their corresponding classical and quantum Hamiltonian dynamics. 
The simple diffusive evolution given by our QW is able to \textit{detect} these two regimes 
which in principle have only been associated to the classical behavior of a free particle inside 
these cavities and its corresponding quantization. 
We also find the presence of a strong scarring phenomena, which in our opinion is a striking result.
This allows to conclude that the simplest generalization of the QW on the line to 2D billiards constitutes a new paradigmatic example of quantum chaos. 

\CB{We should emphasize that our simplest 2D QW detects chaoticity, not through the well-known Wigner-Dyson statistics, but through a combination of weak level repulsion, altered localization properties, and the striking presence of scarring.
The obtained spectra in our calculations are not fully random matrix-like, this suggesting 
that the model captures a different aspect of quantum chaos that is not present in 
Hamiltonian systems.}

Indeed, a deeper theoretical explanation of this behavior is a promising avenue of research,
\CB{but this is a limitation of our work at present}. 
One of the most interesting questions that arises from it is how diffusive systems, 
such as QWs, are able to \textit{detect} chaoticity. 
In particular, which combinations of coin properties and grids do allow for localization 
on POs and which do not. 
Although our work does not permit yet to convincingly conclude it, we foresee that
small perturbations to the Hadamard-like coin keep localization on short orbits,
becoming them deformed. 
On the other hand, the asymmetrical coin case considered breaks all localization, 
except that on the bouncing balls marginally stable orbits. 
Diffusion directions compared to the domain shape and coin combined symmetries should be
relevant, the concrete details of their influence will be studied in the future. 
Moreover, the underlying position grids automatically associated with discrete QWs 
could induce different behaviors when compared to the usual billiard models. 
In fact, a hopping Hamiltonian on lattice billiards has been investigated in \cite{Lattice} 
showing the same spectral properties as the continuum counterparts, 
although this has been done in the open system scenario; 
there new lattice scars were found.
\CB{Then an interesting question arises: Is the connection to latticed Hamiltonian scars direct?}

\CB{To conclude this section, we indicate some relevant possibilities for future research.
In the first place,} more complex situations, such as having two particles inside the domain 
or using other coins, should be considered in the future, since they
would serve as different (more realistic?) models for electrons inside a cavity \cite{Melnikov}. 
\CB{Another interesting point worth studying is} if our \textit{chaotic} QW can 
be used to improve 2D grid searches, as an alternative to the nonlinear QW model studied in \cite{Molfetta}, for instance. 
\CB{Finally, studying the phase space structure of the QW dynamics, with Husimi functions 
for example, should be considered as an interesting aspect in a future research.
Also, the sensitivity of these results to different coin operations and grid geometries is an 
intriguing question.}

\section*{Acknowledgment}
\label{sec:ack} 

This research has been partially supported  by the Spanish Ministry of Science,
Innovation and Universities, Gobierno de Espa\~na under Contract 
No.~2021-122711NB-C21.
Support from CONICET is gratefully acknowledged.

\appendix*\section{1. Details of the QW model}

\CB{This section of the Appendix is devoted to give full details of our QW model.}
Our QW in a rectangular billiard is implemented in the following way. 
We initially apply a SU(2) coin operator of the form
\begin{equation}
 \hat{C}_1 = \left(
   \begin{matrix}
     \cos{\alpha} & \sin{\alpha}  \\
     -\exp{(i \pi/4)} \sin{\alpha} & \exp{(i \pi/4)} \cos{\alpha}
  \end{matrix} \right) ,
\end{equation}
acting on the spin space. 
We take it in this \CB{particular form} in order to break the reflection symmetry 
with respect to the real axis that the spectra \CB{shows} when considering just a rotation.
Then, we proceed with a horizontal displacement operator, acting solely in that direction, 
\CB{which is} given by
\begin{eqnarray}
\hat{W_m} & = & \sum_{0}^{m_{R}-1}|m+1\rangle
\langle m| \otimes |U\rangle \langle U| \nonumber \\
& + & \sum_{1}^{m_{R}}|m-1\rangle \langle m| \otimes
|D\rangle \langle D| + |0,U\rangle \langle 0,D| \nonumber\\
& + & |m_{R},D\rangle \langle m_{R},U|.
\label{eq:3}
\end{eqnarray}
The last two terms correspond to the \CB{effect on the} reflection
at the boundary, i.e.~at $0$ and $m_{R}$, by means of a spin flip.
A proof of \CB{the unitarity of this operator is given}
\CB{in Sect.~2} of this Appendix).
Next, a second coin operator
\begin{equation}
 \hat{C}_2 =
\begin{pmatrix}
	 \cos{\beta} & \sin{\beta}  \\
 -\exp{(i \pi/4)} \sin{\beta} & \exp{(i \pi/4)} \cos{\beta}
  \end{pmatrix}
\end{equation}
is applied.
Finally, the vertical displacement with reflections at $0$ and $n_{U}$,
again acting only on the corresponding direction, is \CB{executed} by
\begin{eqnarray}
\hat{W_n} & = & \sum_{0}^{n_{U}-1}|n+1\rangle \langle n| \otimes |U\rangle \langle U| \nonumber \\
& + & \sum_{1}^{n_{U}}|n-1\rangle \langle n| \otimes
|D\rangle \langle D| + |0,U\rangle \langle 0,D| \nonumber \\
& + & |n_{U},D\rangle \langle n_{U},U|.
\label{eq:5}
\end{eqnarray}
This completes the definition of our evolution operator for one time step.

For the Bunimovich stadium case, we just consider the upper right quarter of the billiard
in order to avoid unwanted spatial symmetries.
The boundary is introduced by modifying Eqs.~(\ref{eq:3}) and (\ref{eq:5}) as follows

\begin{eqnarray}
\hat{W_{m}} & = & \sum_{0}^{w(n)-1}|m+1\rangle
\langle m| \otimes |U\rangle \langle U| \nonumber\\
& + & \sum_{1}^{w(n)}|m-1\rangle \langle m| \otimes
|D\rangle \langle D| + |0,U\rangle \langle 0,D| \nonumber \\
& + & |w(n),D\rangle \langle w(n),U|
\end{eqnarray}
%
\begin{eqnarray}
\hat{W_{n}} & = & \sum_{0}^{f(m) -1}|n+1\rangle
\langle n| \otimes |U\rangle \langle U| \nonumber \\
& + & \sum_{1}^{f(m)}|n-1\rangle \langle n| \otimes
|D\rangle \langle D| + |0,U\rangle \langle 0,D| \nonumber \\
& + & |f(m),D\rangle \langle f(m),U|
\end{eqnarray}

Two shape functions, $f(m)$ and $w(n)$, have been introduced in the two
previous expressions.
Function $f(m)$, which corresponds to the maximum $n$ at each $m$, is given by
\begin{equation}
f(m) = \left\{
\begin{array}{ll}
n_{U} \hspace{2.85cm} \text{if}  \; m \leq m_{C} \\
 \sqrt{n_{U}^{2} - (m - m_{C})^{2}} \quad \text{if} \; m_C \leq m<m_R=2m_C.
 \end{array}
  \right.
  \label{eq:A7}
\end{equation}
Here, $m_{C}$ is the value $m$ at which the circular part of the
boundary begins, while the lower limit is given by $0$, as in the rectangular billiard.
Similarly, function $w(n)$ is the right limit of the displacement
\begin{equation}
  w(n) = m_{C} + \sqrt{n_{U}^{2} - n^{2}} \quad \forall \ n,
  \label{eq:A8}
\end{equation}
with $0$ being the leftmost value, as before. See an
illustration in Fig.~\ref{Fig1}.

\subsection*{2. Unitarity of the evolution operator}

It is worth showing the unitarity of the reflection mechanism at the billiard boundaries. 
Taking the most general vertical displacement: 
\begin{eqnarray}
\hat{W_{n}} & = & \sum_{0}^{f(m)-1} |n+1,U\rangle 
\langle n,U| \nonumber \\ 
& + & \sum_{1}^{f(m)} |n-1,D\rangle \langle n,D| + 
|0,U \rangle \langle 0,D| \nonumber \\
& + & |f(m),D \rangle \langle f(m),U| 
\end{eqnarray}
and,
\begin{eqnarray}
\hat{W}_{n}^{\dagger} & = & \sum_{0}^{f(m)-1} |n,U\rangle
 \langle n+1,U| \nonumber \\
 & + &  \sum_{1}^{f(m)} |n,D\rangle \langle n-1,D| + 
 |0,D \rangle \langle 0,U| \nonumber \\
 & + & |f(m),U \rangle \langle f(m),D| . 
\end{eqnarray}
Hence,  
\begin{eqnarray}
\hat{W}_{n}^{\dagger} \hat{W}_{n} & = & 
\sum_{0}^{f(m)-1} |n,U \rangle \langle n,U| \nonumber\\
& + & \sum_{1}^{f(m)} |n,D \rangle \langle n,D| + 
|0,D \rangle \langle 0,D| \nonumber \\
& + & |f(m),U \rangle \langle f(m),U|, 
\end{eqnarray}
or equivalently
\begin{eqnarray}
\hat{W}_{n}^{\dagger} \hat{W}_{n} & = &
\sum_{0}^{f(m)} |n,U \rangle \langle n,U| 
+ \sum_{0}^{f(m)} |n,D \rangle \langle n,D| \nonumber\\
& = & \sum_{0}^{f(m)} |n \rangle \langle n| \otimes 
( |U \rangle \langle U| + |D \rangle \langle D| ) 
\end{eqnarray}
which is the identity. 
The same happens for the most general horizontal displacement, 
completing the proof.

\end{document}